\documentclass[preprint,12pt]{elsarticle}

\usepackage[T1]{fontenc}
\usepackage[utf8]{inputenc}

\usepackage[english]{babel}

\usepackage[letterpaper,top=2cm,bottom=2cm,left=3cm,right=3cm,marginparwidth=1.75cm]{geometry}

\usepackage{amsmath}
\usepackage{graphicx}
\usepackage{graphviz}
\usepackage[colorlinks=true, allcolors=blue]{hyperref}
\usepackage{algorithm}
\usepackage{algpseudocode}
\usepackage{booktabs}
\usepackage{amssymb}
\usepackage{listingsutf8}

\usepackage{subcaption}
\usepackage{xcolor}
\usepackage{inconsolata}
\usepackage{textcomp}

\lstdefinestyle{shellstyle}{
  basicstyle=\ttfamily\scriptsize\color{black!85},
  backgroundcolor=\color{black!6},
  frame=single,
  rulecolor=\color{black!25},
  framerule=0.4pt,
  xleftmargin=0.6em,
  xrightmargin=0.6em,
  aboveskip=0.4em,
  belowskip=0.4em,
  columns=fullflexible,
  keepspaces=true,
  showstringspaces=false,
  breaklines=true,
  breakatwhitespace=true,
  extendedchars=true,
  moredelim=**[l][\bfseries]{you> },
  literate=
    {á}{{\'a}}1 {é}{{\'e}}1 {í}{{\'i}}1 {ó}{{\'o}}1 {ú}{{\'u}}1
    {Á}{{\'A}}1 {É}{{\'E}}1 {Í}{{\'I}}1 {Ó}{{\'O}}1 {Ú}{{\'U}}1
    {ñ}{{\~n}}1 {Ñ}{{\~N}}1
    {ü}{{\"u}}1 {Ü}{{\"U}}1
    {¿}{{\textquestiondown}}1 {¡}{{\textexclamdown}}1
    {λ}{{$\lambda$}}1
    {model>}{{\textbf{model>}}}6
}

\journal{Simulation Modelling Practice and Theory}

\begin{document} 

\begin{frontmatter}
\title{YAIFS: Yet (not) Another Intelligent Fog Simulator: A Framework for Agent-Driven Computing Continuum Modeling \& Simulation}

\author[]{Isaac Lera}
\ead{isaac.lera@uib.es}

\author[]{Carlos Guerrero}
\ead{carlos.guerrero@uib.es}

\affiliation{organization={University of the Balearic Islands (UIB)},
            addressline={Carretera de Valldemossa, km 7.5},
            city={Palma},
            postcode={07122},
            state={Balearic Islands},
            country={Spain}}

\begin{abstract}
Simulation plays a key role in the design and evaluation of distributed systems, yet it is often treated as a static tool with limited interaction capabilities. In this work, we present \emph{Yet (not) Another Intelligent Fog Simulator} (YAIFS), a substantial evolution of YAFS that redefines simulation as an interactive and service-oriented environment. YAIFS introduces a layered architecture that exposes the simulation through a unified API and a service interface, enabling external entities to observe, control, and modify its execution.
    
A central contribution of this work is the integration of the \emph{Model Context Protocol} (MCP) as a standardized interaction layer between agents and the simulation. Through MCP, heterogeneous agents can access simulation state, invoke actions, and coordinate their behavior using a common set of tools. This abstraction decouples agent logic from simulator internals and supports reproducible and programmable experimentation workflows. 

To illustrate these capabilities, we present two complementary scenarios. With the new framework, we can integrate a large language model (LLM) to provide an interactive assistant that enables users to control simulations through prompts.  Second, we design a multi-agent scenario in which specialized agents monitor system conditions and dynamically adapt application placement decisions. Together, these scenarios highlight how MCP structures agent–simulation interactions and enables adaptive and programmable behavior under dynamic workloads.

The results show that the proposed approach transforms the simulation into an interactive environment that can be systematically explored and controlled by external agents. This perspective opens new directions for AI-driven experimentation in distributed cloud--edge systems, where simulations become programmable entities rather than static evaluation artifacts. The implementation is publicly available at: \url{https://github.com/acsicuib/YAIFS}.
\end{abstract}

  
  \begin{highlights}
    \item Service-oriented simulation for cloud--edge systems
    \item Multi-agent adaptive placement using MCP tools
    \item Natural language interaction via LLM-based agent
    \item Reproducible and programmable simulation workflows
  \end{highlights}

  
\end{frontmatter}

\section{Introduction}

The increasing complexity of cloud–-edge infrastructures has made simulation an essential tool for evaluating system behavior under diverse conditions. However, as these systems grow in scale and heterogeneity, the challenge is no longer limited to accurately modeling their behavior, but also to efficiently exploring a large space of possible configurations and interactions.

Traditional simulation frameworks typically require manual definition and execution of experiments, limiting the ability to systematically analyze alternative scenarios. In this context, when we introduced YAFS in 2019~\cite{yafs_article}, we proposed several innovations that advanced the simulation of cloud continuum architectures. YAFS has had a significant impact on the research community, as reflected by its citation record and by the number of studies that have used it as an experimental framework. In particular, YAFS provided a flexible representation of infrastructure as complex graphs and supported customizable processes that could be invoked during simulation to model dynamic behaviors such as node failures or user mobility. While this approach enabled a high degree of modeling flexibility, it also required users to operate at a relatively low level of abstraction, making the design and execution of experiments a manual and often time-consuming process.

Recent advances in artificial intelligence, particularly large language models (LLMs) and autonomous agents, open new opportunities to rethink how simulation environments are used. Rather than static experimental tools, simulations can be treated as interactive environments that agents can explore, control, and optimize.
For example, users can express high-level goals through prompts such as \emph{“compare two placement strategies minimizing latency”} or \emph{“deploy an application at the edge and evaluate its performance”}, which are then translated into executable simulation workflows.

\begin{figure}[h]
  \centering
  \includegraphics[width=\linewidth]{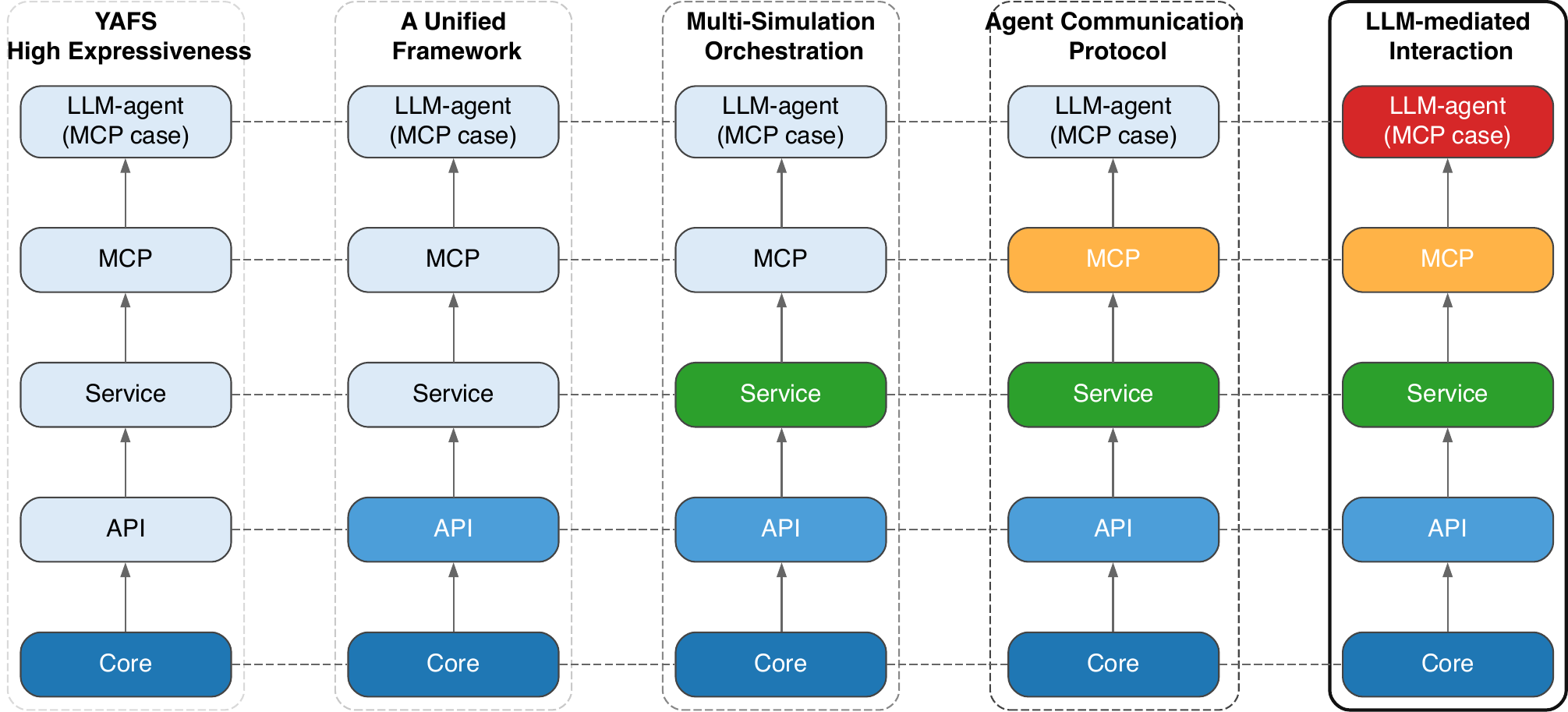}
  \caption{Layered visibility model}
  \label{fig:yaifs_layers}
\end{figure}

In this work, we present a substantial evolution of YAFS, called \emph{Yet (not) Another Intelligent Fog Simulator} (YAIFS). The main novelty lies in a shift in perspective with respect to traditional simulation frameworks, where the simulator is exposed as a layered service architecture, in which each layer defines a different abstraction level and a corresponding distribution of responsibilities between the user and the platform. 
At lower layers, users retain fine-grained control over modeling and execution decisions, which maximizes flexibility at the cost of higher operational responsibility; at higher layers, users can progressively delegate these tasks to the platform. This design mirrors the service hierarchy paradigm in cloud computing, ranging from low-level, user-managed operation to higher-level interfaces that enable more automated workflows.
Importantly, YAIFS itself does not embed decision-making intelligence. Instead, higher-level capabilities can be realized through external agents interacting with the system via the Model Context Protocol (MCP). In this work, we implement an LLM-based agent as a proof of concept, which operates on top of YAIFS to interpret user intents and orchestrate simulation workflows, including scenario modeling, execution, and result analysis.
Figure~\ref{fig:yaifs_layers} summarizes this layered model following a visual logic similar to that used in cloud service hierarchies, progressing from lower-level services with greater user control to higher-level services with increased delegation and automation.


The first layer, \emph{Core}, corresponds to the original version of the simulator (YAFS), where users directly implement simulation logic with full control over the execution. The next layer, \emph{API}, is realized through a unified framework that encapsulates the functionality of the simulator core via an API. At this level, the simulation workflow and parameterization remain under the user's control, but are expressed through higher-level abstractions.

Building on this, we introduce a third layer, the \emph{Simulation as a Service}, that enables the orchestration of multiple simulations. Simulations are no longer treated as static experimental artifacts, but as dynamic entities that can be managed and evolved over time. This layer enables users to monitor, modify, clone, and explore simulations across different execution paths, supporting more complex experimental workflows.

On top of this, we enable interaction with simulations through external agents via the MCP. MCP provides a standardized interface through which agents can access the simulation state and execute actions within the environment. Through this mechanism, agents are able to observe, reason about, and act upon simulations, enabling automated and adaptive experimentation workflows.

Finally, the design paradigm proposed enables interaction through natural language via an additional LLM-mediated layer. In this setting, simulations are mediated by LLM-based agents, allowing users to express only the aspects of the simulation that are relevant to them, while delegating lower-level responsibilities to the system. 

By expressing objectives and experimental intentions at a higher level of abstraction, users can significantly reduce the learning curve associated with complex simulation tools. For instance, a user may specify prompts such as:  \textit{"Compare energy consumption and response time when deploying services closer to the edge versus centralized cloud nodes"} or \textit{"We could reassign these services closer to the cloud cluster to compare the impact on latency with previous results"}. Depending on the capabilities of the underlying model, agents can also support the experimental process by suggesting exploratory actions, such as proposing alternative application placement strategies to evaluate hypothetical scenarios.

In addition, generative AI capabilities enable interactions that go beyond the functionality explicitly implemented in the simulator. For example, given a deployment model produced during a simulation, an agent can generate infrastructure configurations compatible with orchestration tools such as \textit{Terraform} or \textit{Kubernetes}, effectively bridging simulated scenarios and real-world deployments.

The main contributions of this work are summarized as follows:

\begin{itemize}

\item We introduce the concept of \textit{simulation-as-a-service} for cloud--edge ecosystem simulation. YAIFS exposes a programmable API and service interface that enables external control of the simulation lifecycle, while providing built-in metrics and mechanisms to compare system behavior across different temporal windows of the simulation, enabling retrospective and what-if analyses.

\item We integrate the \textit{MCP} as a standardized interface that enables external agents and tools to observe, query, and act upon the simulation environment. As a proof of concept, we implement a CLI-based client that demonstrates this interaction through a single LLM-based agent operating over the MCP interface.

\item We enable the integration of intelligent agents within the experimental workflow, allowing AI-based systems (e.g., LLMs or reinforcement learning agents) to participate in the exploration, control, and evaluation of infrastructure scenarios. This opens the door to more automated, reproducible, and exploratory methodologies for analyzing distributed computing infrastructures.

\item We provide an open-source implementation of YAIFS, including documentation and illustrative scenarios, publicly available in a GitHub repository.

\end{itemize}

Overall, we redefine the role of simulation in the study of distributed systems across the cloud-–edge continuum. Instead of treating simulations as static experimental artifacts executed offline, we frame them as interactive environments that can be observed, controlled, and evolved by external agents. This perspective provides a foundation for AI-driven experimentation in such systems.

The remainder of this paper is organized as follows. Section~\ref{sec:background} discusses the background and related work. Section~\ref{sec:modeling-abstractions} describes the modeling abstractions provided by the simulator for representing cloud–continuum scenarios. Section~\ref{sec:architecture} presents the architecture of YAIFS and its simulation-as-a-service model. Section~\ref{sec:mcp} describes the integration of MCP and illustrates its use through a proof-of-concept implementation, where a LLM-based agent interacts with the simulation across different prompts. Section~\ref{sec:evaluation} presents an illustrative MCP-based multi-agent scenario. Finally, Section~\ref{sec:conclusions} concludes the paper and outlines future research directions.

\section{Related Work}~\label{sec:background}

A wide range of simulation frameworks has been proposed to model and evaluate cloud--edge continuum systems, particularly focusing on application placement, workload offloading, and infrastructure heterogeneity. These simulators provide rich capabilities for representing distributed environments, but they are typically designed to be used through simulator-specific APIs, scripts, or tightly coupled control logic.

iFogSim~\cite{gupta2017ifogsim} is one of the earliest and most widely adopted simulators for fog computing. It enables the modeling of application modules and their placement across cloud and edge resources, allowing researchers to evaluate latency, energy consumption, and resource utilization under different offloading strategies. Building on this foundation, iFogger~\cite{iFogger2024} extends simulation capabilities within the OMNeT++ ecosystem, emphasizing topology-aware placement and dynamic workload routing using complex-network measures.

PureEdgeSim~\cite{mechalikh2021pureedgesim} provides a flexible framework for evaluating orchestration strategies across cloud, fog, and edge infrastructures. It includes support for containerized deployments, mobility-aware scenarios, and different placement policies. SatEdgeSim~\cite{wei2020satedgesim} further extends PureEdgeSim to model satellite–edge environments, enabling the study of deployment and offloading strategies in scenarios where satellite links introduce additional latency and energy considerations.

DISSECT-CF-Fog~\cite{DISSECT2022} integrates cloud simulation with fog computing extensions, incorporating TOSCA-based deployment descriptions and predictive models to support dynamic placement and capacity-aware orchestration. ENIGMA~\cite{del_pozo_punal_2021_10953027} focuses on large-scale, location-aware simulations involving mobile devices, supporting experiments on workload distribution and offloading in highly dynamic environments. More recently, EdgeAISim~\cite{NANDHAKUMAR2024100939} has been proposed to study AI-driven workloads at the edge, and ECLYPSE~\cite{Massa_2026} combines simulation and emulation to model the cloud--edge continuum with a focus on adaptability and reproducibility.

Comprehensive surveys such as~\cite{MARKUS2020102042,mechalikh2025quality} highlight the maturity of this ecosystem, typically comparing simulators in terms of modeling capabilities, scalability, realism, and usability. These studies show that existing tools are well-suited for evaluating placement strategies and system behavior under diverse conditions.

However, despite their strengths, most existing simulators share a common interaction paradigm: they are primarily designed to be programmed directly by the user, with experimental logic embedded in scripts or simulator-specific APIs. As a result, the interaction between external decision-making entities and the simulation is often ad hoc and tightly coupled to the implementation.

In contrast, our work focuses on redefining this interaction model. YAIFS exposes simulation as a service and introduces the MCP as a standardized interaction layer. This approach enables external agents, ranging from LLM-based assistants to multi-agent systems, to observe, query, and act upon the simulation through a unified set of tools.

\section{Modeling}~\label{sec:modeling-abstractions}

YAIFS builds upon the modeling perspective introduced in YAFS for representing distributed infrastructures in the cloud continuum. In this approach, the infrastructure is described as a graph-based topology of computational resources, applications are modeled as directed acyclic graphs (DAGs) representing service dependencies, and users are represented as request-generating modules located at nodes of the topology.

Request routing and application placement are controlled through configurable selection mechanisms that determine how messages traverse the infrastructure and where application components are deployed. These mechanisms can be customized to represent different routing policies or placement strategies.

To support the new interaction layers, we define a consistent set of modeling abstractions that can be safely manipulated through these interfaces. The following subsections describe the main entities of the simulation model and the metrics derived from simulation execution.

\subsection{Infrastructure design}

The computational infrastructure is modeled as a set of interconnected clusters representing resources distributed across the cloud–edge continuum. Each cluster groups a collection of nodes that share local connectivity and administrative boundaries.

Clusters are typically associated with conceptual layers of the cloud-–edge infrastructure. In our examples, we adopt a common terminology where \textit{CDC} (Cloud Data Center), \textit{MEC} (Multi-access Edge Computing), and \textit{EDC} (Edge Data Center) represent different levels of proximity to end users. However, these roles are not enforced by the simulator, and users remain free to define arbitrary cluster types depending on the experimental scenario.

Within each cluster, nodes represent individual computational resources. For modeling simplicity, nodes belonging to the same cluster are assumed to have full internal connectivity. Nodes may have different roles, such as \textit{control-plane} nodes responsible for maintaining connectivity with other clusters and \textit{worker} nodes hosting application components.

Clusters are connected through physical links that define the inter-cluster network topology. These links include attributes such as geographical distance and target latency, which are used to estimate propagation delays and support placement decisions based on network characteristics.

In addition, we introduce the concept of \textit{region}, which allows grouping clusters according to geographical or administrative boundaries. This abstraction enables the computation of traffic-related metrics such as ingress and egress costs between regions.

\begin{figure}[h]
  \centering
  \includegraphics[width=0.5\textwidth]{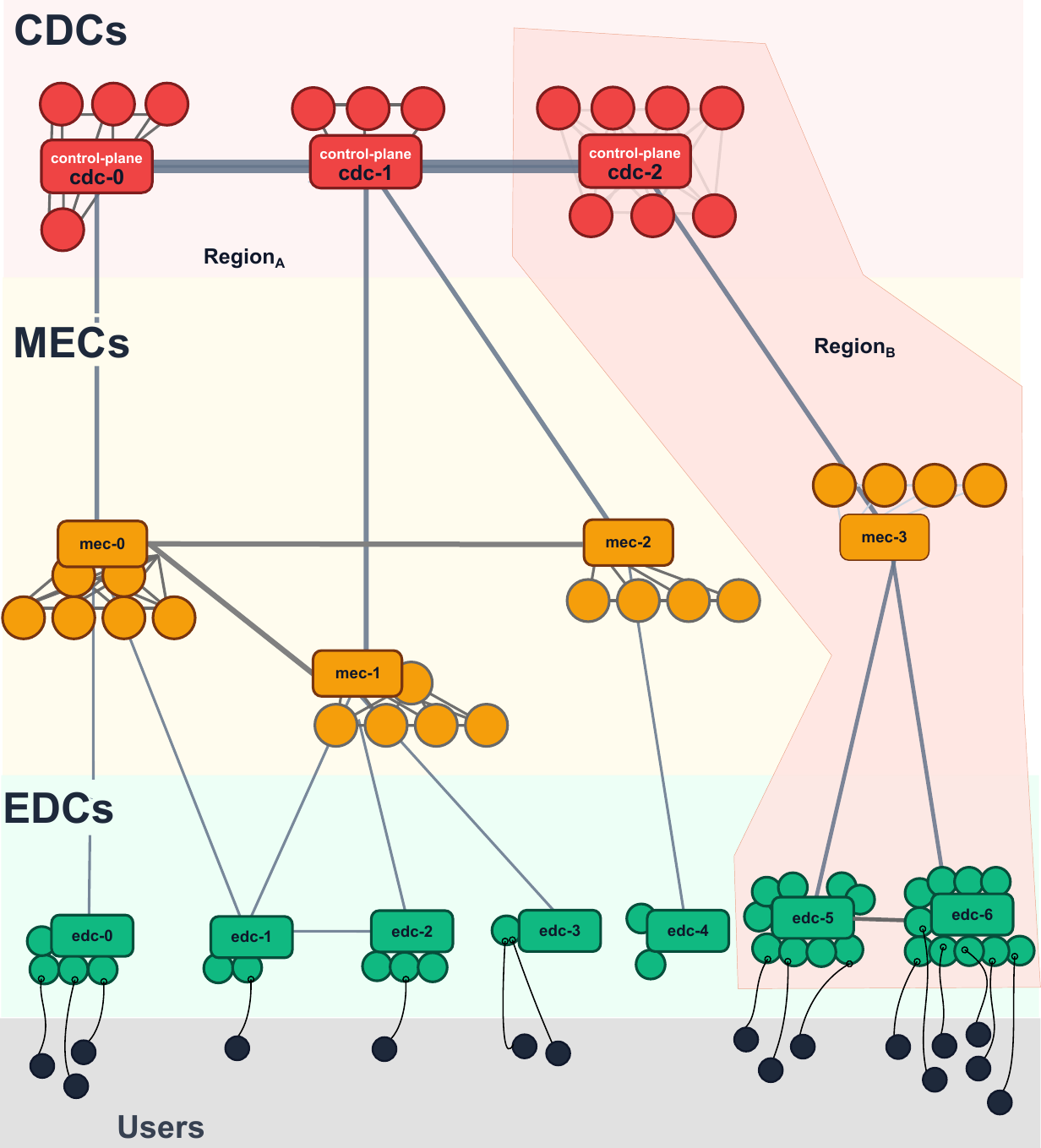}
  \caption{Example topology of a cloud–continuum infrastructure composed of 3 \textit{CDC} clusters, 4 \textit{MEC} clusters, and 7 \textit{EDC} clusters, each containing multiple computational nodes.}
  \label{fig:topology}
\end{figure}

Figure~\ref{fig:topology} shows an example topology of a cloud–continuum infrastructure. Listing~\ref{lst:topology_example} presents a simplified excerpt of the JSON configuration used to define such infrastructures in YAIFS.

\begin{lstlisting}[style=shellstyle,caption={Example infrastructure configuration},label={lst:topology_example}]
  "clusters": [{
      "name": "cdc-0",
      "role": "CDC",
      "region": "region-a",
      "nodes": [{
          "name": "cdc-0-control-plane",
          "role": "control-plane",
          "capacity": {"cpu": "100.0", "memory": "100000m"}
      }]
  }]
  \end{lstlisting}

  Applications are modeled following an ETSI-like description of network services. In this model, an application is composed of a set of Virtual Network Functions (VNFs) executed in a specific order to resolve a service request. The service logic is therefore represented as a VNF chain, where each function processes the request and forwards it to the next stage.

  This representation remains compatible with the original service model introduced in YAFS, where applications are defined as directed processing flows across multiple modules. The main extension is the ability to explicitly define a backward response path from the final VNF to the user that initiated the request. This mechanism allows modeling scenarios in which responses must traverse the network back to the originating user.

User behavior follows the model originally introduced in YAFS. Users are represented as request-generating entities associated with a specific application. Each user specifies the invoked application, the distribution governing request generation, and the message sent to the first VNF of the service chain. This abstraction allows the simulator to represent different workload patterns by configuring the statistical properties of request generation.

We also support the definition of custom processes that model dynamic events occurring during the simulation. These processes capture phenomena such as user mobility, node failures, node recoveries, or dynamic workload changes. They can be defined either as user-provided functions executed during simulation events or through structured descriptors interpreted by the simulator. 

For example, the following process models user mobility by randomly moving users between candidate nodes every 30 time units.

\begin{lstlisting}[style=shellstyle,caption={Example process definition for user mobility},label={lst:process_example}]
{
  "name": "user-mobility-ar",
  "kind": "user_mobility_random",
  "enabled": true,
  "distribution": {
    "type": "deterministic",
    "time": 30
  },
  "params": {
    "app_ref": "Augmented Reality (AR)",
    "nodes": ["mec-0-worker-1", "edc-1-worker-1", "edc-2-worker-1"],
    "create_probability": 0.10,
    "move_probability": 0.80
  }}
\end{lstlisting}


Application placement corresponds to the assignment of VNFs to nodes in the infrastructure topology. Placements may be defined explicitly or generated dynamically through placement strategies implemented as simulation processes.

By default, message routing between VNFs follows the shortest path between the nodes hosting the corresponding functions. This default behavior can be overridden by custom routing policies that take into account additional criteria such as network latency, link cost, or infrastructure constraints.

\subsection{Metrics design}

We group the metrics provided by the simulator into several categories that capture service performance, infrastructure utilization, network behavior, and operational costs.

\begin{itemize}

  \item \textbf{Application-level metrics.} These metrics describe the performance experienced by service requests as they traverse the VNF chain of an application. They include indicators such as the total number of processed requests, the number of successful and unsuccessful responses, and latency statistics (mean response time, percentiles such as p50 and p95, and maximum response time). The response time is further decomposed into its main components, including network transmission time, processing time at each VNF, and waiting time caused by resource contention.

  \item \textbf{Infrastructure utilization metrics.} These metrics describe how computational resources are used during the simulation. They include node utilization, cluster utilization, and the distribution of users across nodes. Such indicators help evaluate how efficiently the infrastructure is used and how workloads are distributed across the cloud–-edge hierarchy.

  \item \textbf{Network metrics.} Network-related indicators characterize how requests traverse the infrastructure topology. These include link utilization, the number of hops per request, and aggregated measurements of network congestion. The simulator also estimates the geographical distance traveled by requests and application responses, allowing the analysis of propagation effects and routing efficiency.

  \item \textbf{Cost and resource consumption metrics.} To support economic analysis of deployment strategies, the simulator computes cost-related indicators associated with application placement and network traffic. These include placement cost, egress traffic cost between regions, and the total cost of a deployment configuration.

  \item \textbf{Runtime monitoring metrics.} Lightweight runtime snapshots summarize the current state of the simulation, including entity-level counters and connectivity anomalies such as unreachable links. These metrics are useful for monitoring long-running simulations or for agent-driven experimentation workflows.

\end{itemize}

All these metrics are accessible through the different interaction layers of the system. At the API level they are computed directly from simulation traces, while the service-oriented interface exposes them as simulation-scoped operations. Through the MCP transport layer, the same metrics can be retrieved by external tools or autonomous agents, enabling automated analysis and comparison of experimental scenarios.

\section{Architecture of YAIFS}~\label{sec:architecture}



We organize the system architecture into three main layers: \textit{Core}, \textit{API}, and \textit{Service Interface}. Each layer exposes a different level of abstraction for interacting with the simulator, allowing users and external agents to operate the system according to their desired level of control.

\subsection{Core layer}

The \textit{Core} layer provides the fundamental primitives of the simulator and implements the discrete-event execution engine inherited from YAFS. Internally, it is built on a \texttt{simpy} environment that orchestrates application, placement, population, and custom processes over simulation time. The \textit{Core} also implements a network-level message pipeline where transmission delays are computed from link bandwidth and propagation attributes, while dynamic link occupancy is tracked to model contention effects. In addition, it maintains explicit deployment mappings between DES processes, application modules, and topology nodes, and integrates path-selection and failure-handling mechanisms to support rerouting and loss accounting under topology changes. Finally, the layer records fine-grained node and link metrics during execution. The design intentionally preserves compatibility with the original architecture while introducing minimal extensions required by YAIFS.

\subsection{API layer}

On top of this foundation, the \textit{API} layer provides higher-level abstractions for modeling applications, users, and infrastructure elements while orchestrating the execution of the underlying simulation engine. This layer encapsulates simulation lifecycle management (e.g., initialization, execution control, pause, resume, and termination) and exposes a set of standardized domain operations for interacting with the simulated environment.

Through this interface, users can create and modify infrastructure elements such as clusters and nodes, register applications, instantiate users, and dynamically deploy, migrate, replicate, or remove application components during runtime. The API also supports the definition of dynamic processes-including mobility patterns, failures, recoveries, and service relocations-while maintaining internal consistency mechanisms such as routing cache updates after topology changes.

By encapsulating these mechanisms, the API reduces the complexity of interacting with the discrete-event engine while providing a structured interface that can be safely accessed by external tools and autonomous agents.

\subsection{Service Interface layer}
Finally, the \textit{Service Interface} layer exposes the simulator as a service-oriented environment in which simulations are treated as managed, long-lived entities. Through this interface, simulations can be created, initialized, executed, paused, resumed, or terminated, enabling external tools and agents to orchestrate the full lifecycle of an experiment.

In the current implementation, each simulation maintains explicit lifecycle states (e.g., created, initialized, running, paused, stopped, or failed) and supports controlled execution through scheduled runtime windows. To ensure consistency, runtime modifications to topology, applications, users, or processes are restricted to actionable non-running states.

The service layer also enables branch-oriented experimentation through safe \textit{fork} operation that duplicate the full simulation state once execution is paused or idle. This allows researchers to explore alternative configurations-such as different placement policies, workload conditions, or failure scenarios-while preserving the lineage between the original and derived simulations.

Additionally, the interface exposes introspection and metric endpoints, including state snapshots, time-windowed application metrics, network statistics, and deployed entities. These capabilities enable external clients and autonomous agents to coordinate reproducible, comparative, and exploratory experimentation workflows.



Table~\ref{tab:primitive-mapping} compares representative primitives across the three access layers, showing which operation should be invoked depending on whether the interaction occurs at one specific interface.

\begin{table}[ht]
  \scriptsize
  \centering
  \caption{Small subset of primitives and their mapping across \textit{core}, \textit{api}, and \textit{service interface} layers.}
  \label{tab:primitive-mapping}
  \begin{tabular}{p{3.2cm}|p{3.9cm}|p{3.4cm}|p{3.4cm}}
  \hline
  \textbf{Primitive} \newline Object: & \textbf{Core} \newline (none) & \textbf{API} \newline Simulation. & \textbf{Service interface} \newline Service. \\
  \hline
  Create simulation & Object construction and setup: \texttt{Sim(...)}  & \texttt{Simulation(\dots..)} & \texttt{.create\_simulation(...)} \\\hline
  Create one user/source & \texttt{Sim.deploy\_source(app\_name, id\_node, msg, distribution)} & \texttt{.create\_users(...)} & \texttt{.create\_users(sim\_id, definition, nodes=...)} \\\hline
  Create users in batch & Repeated: \texttt{deploy\_source(...)} calls & \texttt{.create\_users(definition, nodes=...)} & \texttt{.create\_users(...)} \\\hline
  Register custom process & \texttt{Sim.deploy\_monitor(name, function, distribution, ...)} & \texttt{.register\_process(...)}  & \texttt{.create\_process(sim\_id, definition)} \\\hline
  Run simulation window & \texttt{Sim.run(until=...)} & \texttt{.run\_for(duration, step=...)} & \texttt{.run\_for(sim\_id, ...)} \\\hline
  Pause / resume execution & Controlled at process level & \texttt{.pause()}\newline \texttt{.resume()} & \texttt{.pause(sim\_id)}\newline \texttt{.resume(sim\_id)} \\\hline
  Fork simulation state & Manual state management & \texttt{.fork(...)} & \texttt{.fork(sim\_id, ...)} \\\hline
  Wait until idle state & N/A (single-engine execution primitive) & N/A (no orchestration wait primitive) & \texttt{.wait\_until\_idle(sim\_id, ...)} \\\hline
  Application metrics summary & N/A (raw metrics collection in engine) & N/A (no aggregated service-level summary) & \texttt{.get\_app\_metrics(sim\_id, ...)} \\\hline
  \hline
  \end{tabular}
\end{table}

\section{MCP Integration \& LLM scenarios}~\label{sec:mcp}

To enable interaction between external agents and the simulation environment, we integrate the MCP as a standardized communication layer. Through MCP, agents can discover available operations, access the simulation state, and perform actions through a common protocol interface, enabling interoperable and reproducible experimentation workflows.

MCP acts as a bridge between the simulation service and external agent runtimes. Instead of interacting directly with internal simulator APIs, agents communicate through MCP tools and resources that expose the main capabilities of the environment. Through this interface, agents can create and control simulations, modify experimental conditions, trigger system events, and retrieve metrics describing the state and performance of the infrastructure.

This integration allows the simulation to be treated as an interactive environment that can be observed and manipulated by autonomous agents. In particular, MCP enables LLM-based agents or reinforcement learning systems to reason about the simulation state and coordinate experimental actions through a standardized interface.

To illustrate agent interaction with the simulation environment, we implemented a simple command line interface (CLI) that communicates with the MCP server. Through this interface, user requests expressed in natural language are interpreted by a LLM and translated into MCP tool calls that operate on the simulator. In the next scenarios, we used the OpenAI GPT-5 model to perform this interpretation. In addition, the application provided in the repository logs all interactions, including user prompts, model responses, tool calls, and the configurations generated to represent the requested or inferred scenarios. It also records the execution traces of the simulations, enabling full traceability and reproducibility of the experimental workflows. This logging capability is particularly relevant for analyzing agent behavior and understanding how high-level requests are decomposed into sequences of simulation operations.

The first interaction illustrates how the agent interprets the capabilities available through the MCP tools exposed by the simulator. When the user asks what the system can do, the language model summarizes the set of available operations and presents them as high-level functionalities that can be executed through the simulation interface. As shown in Listing~\ref{lst:conversation_create_0}, the agent describes tasks such as creating simulations, deploying applications, modifying placements, managing user traffic, running experiments, and collecting metrics. The example also highlights another relevant property of the interaction: the agent naturally adapts its responses to the language used by the user. In this case, after the initial query in English, the user asks the same question in Spanish and the agent replies accordingly in Spanish.

\begin{lstlisting}[style=shellstyle,caption={Agent interpretation of the available YAIFS tools and capabilities.},label={lst:conversation_create_0}]
you> what can you do?
model> I can help you **set up, run, and analyze YAIFS/YAFS simulations**. Concretely, I can:
- **Create simulations** (from a scenario directory or from a generated configuration: clusters/nodes, topology, services, users, placements).
- **Inspect state** of a simulation (nodes, clusters, deployed apps/VNFs, processes).
- **Deploy/replicate VNFs and applications** (and optionally choose placement strategies/allowed nodes).
- **Move or remove VNFs** to create **different placement variants**.
- **Manage user traffic** (add/remove users, move users to nodes, adjust their traffic rate λ).
- **Run the simulation** (run for a duration, schedule, pause/resume, stop, destroy).
- **Collect metrics** and compare outcomes:
  - application-level metrics (latency/critical path, etc.)
  - network-level metrics (link/egress cost/utilization depending on scenario)

If you tell me your goal (e.g., *compare two placements*, *minimize latency*, *minimize cost*, etc.) and provide a `simulation_id` (or let me create one), I'll propose the experiment and execute it.

you> ¿qué puedes hacer?
model> Puedo ayudarte a diseñar, ejecutar y analizar simulaciones YAIFS/YAFS. En concreto, puedo:
...
\end{lstlisting}

The next interaction illustrates how a user can create a simulation scenario using natural language. The agent interprets the request and generates the corresponding infrastructure through the available MCP tools.
As shown in Listing~\ref{lst:conversation_create_1}, the simulation is created and a unique identifier is returned, allowing the user or the agent to continue configuring the scenario before executing the experiment. In addition to natural language descriptions, YAIFS also supports defining scenarios through structured configuration files (e.g., JSON) that can be given to the MCP client.

\begin{lstlisting}[style=shellstyle,caption={Agent-driven creation of a simulation infrastructure},label={lst:conversation_create_1}]
you> Create a simulation with three clusters, each with 10 nodes.
[03/25/26 17:47:16] INFO     Processing request of type CallToolRequest  server.py:720
[03/25/26 17:47:17] INFO     Processing request of type CallToolRequest  server.py:720
model> Done. I created a new YAIFS/YAFS simulation with **3 clusters** and **10 nodes per cluster** (**30 nodes total**).

- **simulation_id:** `sim-35a3ffb7`
- **name:** `clusters3_nodes10`
- **status:** created (not started yet)

If you want, I can now deploy the default applications/VNFs and run it, or we can create two different placement variants and compare them.
you> 
\end{lstlisting}

The next interaction~\ref{lst:conversation_made_simulation} illustrates how a high-level user request is decomposed into a sequence of MCP tool calls that collectively implement the experiment. The internal sequence of MCP tool calls generated by the agent is as follows: 
{\small
\texttt{crea\-te\_de\-fault\_si\-mu\-la\-tion}, 
\texttt{list\_simulations}, 
\texttt{crea\-te\_si\-mu\-la\-tion\_appli\-ca\-tion}, 
\texttt{crea\-te\_si\-mu\-la\-tion\_appli\-ca\-tion}, 
\texttt{get\_si\-mu\-la\-tion\_state}, 
\texttt{list\_si\-mu\-la\-tion\_pro\-cesses}, 
\texttt{list\_si\-mu\-la\-tion\_de\-plo\-yed\_app\-li\-ca\-tions}, 
\texttt{list\_\-si\-mu\-la\-tion\_appli\-cation\_vnfs}, 
\texttt{list\_\-si\-mu\-la\-tion\_appli\-cation\_vnfs}, 
\texttt{list\_\-si\-mu\-la\-tion\_users}, 
\texttt{run\_si\-mu\-la\-tion\_for}, 
\texttt{wait\_si\-mu\-la\-tion\_until\_ready}, and 
\texttt{get\_si\-mu\-la\-tion\-\_app\-li\-cation\_me\-trics}.
}
This sequence reveals that the agent explores the available tool space rather than following a strictly optimal execution path. In particular, some operations are redundantly invoked (e.g., {\small \texttt{create\_si\-mu\-la\-tion\_appli\-ca\-tion}} and {\small \texttt{list\_si\-mu\-la\-tion\_appli\-cation\_vnfs}}), while others correspond to exploratory inspection steps (e.g., {\small \texttt{list\_si\-mu\-la\-tion\_pro\-cesses}}, {\small \texttt{list\_si\-mu\-la\-tion\_users}}) that are not strictly required to fulfill the user request.

A more optimal sequence would consist of a reduced set of operations: (i) {\small \texttt{create\_de\-fault\_si\-mu\-la\-tion}}, (ii) {\small \texttt{run\_si\-mu\-la\-tion\_for}}, (iii) {\small \texttt{wait\_si\-mu\-la\-tion\_until\_ready}}, and (iv) {\small \texttt{get\_si\-mu\-la\-tion\_appli\-cation\_me\-trics}}.

\begin{lstlisting}[style=shellstyle,caption={Agent planning: tool call decomposition},label={lst:conversation_made_simulation}]
you> Create a baseline simulation, deploy an application and a user randomly, and run it for 1000 time units. Then analyze the response times.
(...)
**Response time analysis (per application):**
**Augmented Reality (AR):**
- Requests: 8 (successful: 8 / failed: 0)
- response_mean: 69.90
- response_p50: 69.90
- response_p95: 74.90
- response_max: 74.90
- processing_mean: 0.90
- waiting_mean: 0.0

**mIoTs:** 
- Requests: 4 (successful: 4 / failed: 0)
- response_mean: 35.60
- response_p50: 35.60
- response_p95: 35.60
- response_max: 35.60
- processing_mean: 0.30
- waiting_mean: 0.0
(...)
\end{lstlisting}

The final interaction (Listing~\ref{lst:conversation_create_2}) illustrates a higher-level request posed by the user: comparing two alternative placement strategies. In this case, the simulator layers does not provide a single primitive operation to perform this task. Instead, the agent interprets the request and plans a sequence of actions using the available MCP tools, including generating alternative placements, executing the corresponding simulations, and comparing the resulting metrics. The agent first proposes possible placement variants and then requests the necessary information (e.g., a simulation identifier or scenario definition) to execute the comparison. This example highlights how the agent can translate abstract experimental goals expressed in natural language into concrete simulation workflows.

\begin{lstlisting}[style=shellstyle,caption={Agent planning skills},label={lst:conversation_create_2}]
you> Are you able to suggest two different placements and compare them with each other?
model> Yes if you already have a YAIFS/YAFS simulation (or a scenario I can load), I can:

  1) Propose **two alternative placement strategies/topologies** (e.g., "spread across clusters" vs "pack into fewer clusters", or "pin to allowed nodes" sets).
  2) Apply them in the simulator by **deploying/moving VNF instances** accordingly.
  3) **Compare results** using simulation metrics (application/network metrics, critical path/latency, throughput, link usage, etc.).
  
  To do this concretely, I need one of the following from you:
  
  - **Option A:** Tell me the **simulation_id** you want to compare (and which apps/VNFs are involved).
  - **Option B:** Share the **scenario directory path** (the one containing `topology.json`, `services.json`, `placements.json`, `users.json`, etc.), or let me create/load 
  a default one.
  
  Also, for the two placements, what do you mean by "different"?
  - By **cluster** (e.g., Cluster A/B vs A/C)?
  - By **node constraints** (e.g., "only edge nodes" vs "only core nodes")?
  - By **objective** (minimize latency vs minimize cost vs maximize reliability)?
  
  If you reply with a simulation_id (or scenario path) + your intended placement difference, I'll run the two placements and produce a side-by-side comparison.
you>
\end{lstlisting}

\section{Experimental MCP-based Scenario and Evaluation}
\label{sec:evaluation}

To evaluate the capabilities of the proposed MCP-based multi-agent approach, we design a dynamic cloud--edge scenario that introduces workload variations and localized demand hotspots. The goal is not only to analyze how different placement strategies behave under changing conditions, but also to illustrate how agent-based orchestration can adapt the system through structured interaction with the simulation. We consider three placement strategies:

\begin{itemize}
  \item \textbf{Random placement}: applications are initially deployed randomly across the infrastructure. A fixed number of initial replicas (20) is created, and users are assigned to shared instances in a round-robin fashion.
  
  \item \textbf{Greedy placement}: application instances are deployed as close as possible to users, minimizing network distance. A new instance of the application is created for each user, resulting in no sharing of VNF chains.
  
  \item \textbf{Multi-agent MCP placement}: the simulation evolves dynamically in discrete time windows, where agents monitor system metrics and adapt placement decisions during execution.
\end{itemize}

The random and greedy strategies are designed as interpretable baselines. The greedy approach is expected to provide lower latency, as each user is served by a dedicated application instance deployed at the edge. However, this comes at the cost of a significantly higher number of replicas and increased deployment cost due to the use of MEC nodes. In contrast, the random strategy is expected to exhibit higher average latency, as VNF chains are distributed across the infrastructure, potentially increasing network congestion. The multi-agent approach is expected to achieve an intermediate behavior, balancing performance and resource usage by dynamically adapting deployments to current system conditions.

The remainder of this section is organized as follows. We first describe the scenario configuration, then detail the multi-agent design, followed by the experimental results. Finally, we discuss the role of MCP in structuring the interaction between agents and the simulation, which constitutes the main objective of this experimental study.

\subsection{Infrastructure Description}
The topology spans \textit{CDC}, \textit{EDC}, and \textit{MEC} clusters, comprising 3, 10, and 27 clusters, respectively, with a total of 214 nodes distributed non-uniformly across them. User requests are generated at \textit{MEC} nodes, while application services can be deployed across the entire infrastructure. Applications are modeled as service chains with different latency requirements. The evaluated workload includes three representative applications: \emph{Perception Pipeline}, \emph{Coordination Pipeline}, and \emph{Telemetry Monitoring}, which differ in complexity and latency sensitivity as summarized in Table~\ref{tab:applications}.

\begin{table}[t]
  \scriptsize
  \centering
  \caption{Application characteristics in the evaluated scenario}
  \label{tab:applications}
  \begin{tabular}{lccc}
  \hline
  \textbf{Application} & \textbf{\# VNFs} & \textbf{\# Messages} & \textbf{Latency Requirement (simulation units)} \\
  \hline
  Perception Pipeline     & 3 & 4 & 120 \\
  Coordination Pipeline   & 2 & 3 & 75 \\
  Telemetry Monitoring    & 1 & 2 & 50 \\
  \hline
  \end{tabular}
  \end{table}

\subsection{Multi-Agent MCP Placement Strategy Design}

We focus our description on the MCP-based multi-agent approach. The random and greedy placement strategies are used as baselines and are briefly discussed in the results comparison. The proposed strategy relies on the interleaved execution of two agents throughout the simulation. This design illustrates the role of the MCP layer while keeping the system simple and interpretable.

The simulation follows a window-based execution model:

\begin{enumerate}
  \item The \textit{Monitoring Agent} observes the simulation over a time window $W_k$ and produces a system snapshot $S_k$.
  \item The \textit{Placement Agent} selects a strategy based on $S_k$.
  \item A set of candidate placement actions is generated.
  \item Up to a maximum of $B$ actions are executed ($B = \texttt{action\_bud\-get\_per\_win\-dow}$).
  \item The simulation advances to the next window or terminates when the predefined simulation horizon is reached.
\end{enumerate}

The following subsections describe the behavior of each agent in detail.

\subsubsection{Monitoring Agent}

The \textit{Monitoring Agent} observes the state of the simulation at the end of each actionable time window and produces a structured snapshot that is later consumed by the \textit{Placement Agent}. It does not modify the system state; rather, its role is to transform low-level metrics into a compact representation of the current operational condition of the infrastructure.

Let the simulation be partitioned into actionable windows
\[
W_k = [t_k, t_{k+1}), \qquad |W_k| = t_{k+1} - t_k,
\]
where $|W_k|$ denotes the duration of the $k$-th window. At the end of each window, the monitoring agent produces a state summary.
\[
S_k = \Big(
\mathcal{T}_k,
\mathcal{R}_k,
\mathcal{U}^{\text{node}}_k,\,
\mathcal{O}_k,\,
\mathcal{U}^{\text{link}}_k,\,
\mathcal{C}_k
\mathcal{P}_k, 
\Big),
\]



These are computed over the current window as follows.

\paragraph{Response\_p95 ($\mathcal{T}_k$)}
For each application $a$, let $\mathcal{R}(a,W_k)$ be the set of completed service requests of application $a$ whose end time belongs to $W_k$. For each request $r \in \mathcal{R}(a,W_k)$, let $\mathrm{resp}(r)$ denote its end-to-end service response time, including return messages when applicable. The latency indicator used by the agent is the empirical 95th percentile:
\[
\mathrm{response\_p95}(a,W_k)
=
Q_{0.95}\Big(\{\,\mathrm{resp}(r)\;|\; r \in \mathcal{R}(a,W_k)\,\}\Big),
\]
where $Q_{0.95}$ denotes the empirical 95th percentile. An application is marked as degraded if
\[
\mathrm{response\_p95}(a,W_k) > \mathrm{latency\_requirement(a)}
\]

\paragraph{Unsuccessful requests ($\mathcal{R}_k$)}
This indicator captures delivery failures, incomplete service chains, and response losses. Let $\mathrm{requests\_total}(a,W_k)$ be the number of requests emitted for application $a$ during $W_k$, and let $\mathrm{requests\_successful}(a,W_k)$ be the number of requests that complete at least one valid service path within the same observation horizon. The number of unsuccessful requests is defined as

\[
\begin{split}
\mathrm{requests\_unsuccessful}(a,W_k) = & \max\Big( \mathrm{requests\_total}(a,W_k) \\
& - \mathrm{requests\_successful}(a,W_k),\,0 \Big).
\end{split}
\]

An application-level incident is raised whenever
\[
\mathrm{requests\_unsuccessful}(a,W_k) > 0.
\]

\paragraph{Node utilization ($\mathcal{U}^{\mathrm{node}}_k$) and overloaded nodes ($\mathcal{O}_k$)}

For each node $n$, the monitoring agent estimates the effective computational load from computation events recorded during the window. Let $\mathcal{M}(n,W_k)$ be the set of computation events executed at node $n$ whose completion time belongs to $W_k$. For each event $m \in \mathcal{M}(n,W_k)$, let
\[
\mathrm{service\_time}(m) = t_{\mathrm{out}}(m) - t_{\mathrm{in}}(m)
\]
be the service time of the event. The window-based node utilization is then defined as
\[
\mathrm{node\_utilization}(n,W_k)
=
\frac{\sum_{m \in \mathcal{M}(n,W_k)} \mathrm{service\_time}(m)}
{|W_k|}.
\]
A node is considered overloaded if
\[
\mathrm{node\_utilization}(n,W_k) > \theta_{\text{node}},
\]
where $\theta_{\text{node}}$ is the node utilization threshold. In the current configuration, $\theta_{\text{node}} = 0.08$. We empirically set this threshold to a relatively low value in order to induce a diverse set of actions, since the workload is highly distributed across a large number of nodes. If this condition is maintained for a number of consecutive windows greater than or equal to $\omega_{\text{overload}}$, the node is inserted into the overloaded set $\mathcal{O}_k$. In the current scenario, $\omega_{\text{overload}} = 1$.

\paragraph{Link utilization ($\mathcal{U}^{\text{link}}_k$) and Congested links ($\mathcal{C}_k$)}
For each link $e$, let $\mathcal{T}(e,W_k)$ be the set of transmitted messages traversing $e$ during $W_k$. If $\mathrm{size}(m)$ denotes the size of message $m$, the average bandwidth used over the window is
\[
\mathrm{bandwidth\_used}(e,W_k)
=
\frac{\sum_{m \in \mathcal{T}(e,W_k)} \mathrm{size}(m)}
{|W_k|}.
\]
Let $\mathrm{bandwidth\_available}(e)$ be the nominal capacity of link $e$. The link utilization is computed as
\[
\mathrm{link\_utilization}(e,W_k)
=
\frac{\mathrm{bandwidth\_used}(e,W_k)}
{\mathrm{bandwidth\_available}(e)}.
\]
A link is marked as congested if
\[
\mathrm{link\_utilization}(e,W_k) > \theta_{\text{link}},
\]
where $\theta_{\text{link}}$ is the congestion threshold. In the current scenario, $\theta_{\text{link}} = 0.5$. Congested links are collected into the set $\mathcal{C}_k$.

\paragraph{Placement cost ($\mathcal{P}_k$)}
The monitoring agent also observes the aggregated placement cost associated with the currently deployed services. If the aggregated placement cost exceeds the configured budget $\theta_{\text{cost}}$, the placement agent may activate a cost-oriented strategy. In the current configuration, $\theta_{\text{cost}} = 8.7$.

Formally, for an application $a$, the placement cost is defined as:
\[
\mathrm{placement\_cost}(a)
=
\sum_{d \in \mathcal{D}(a)} \mathrm{cost}\big(\mathrm{node}(d)\big),
\]
where $\mathcal{D}(a)$ denotes the set of active VNF deployments for application $a$, and $\mathrm{node}(d)$ is the node hosting deployment $d$.  In our scenario, node costs reflect the proximity to end users: \textit{CDC} nodes have a cost of $0.01$, \textit{EDC} nodes $0.06$, and \textit{MEC} nodes $0.30$. 

\subsubsection{Placement Agent}

The \textit{Placement Agent} operates at the end of each actionable time window $W_k$, consuming the system snapshot $S_k$ produced by the \textit{Monitoring Agent}. The agent selects a placement strategy $\sigma_k$ through a hierarchical rule-based policy:
\[
\sigma_k =
\begin{cases}
\texttt{Cost}, & \text{if } \mathcal{P}_k > \theta_{\text{cost}}, \\
\texttt{Overload}, & \text{if } \mathcal{O}_k \neq \emptyset, \\
\texttt{Congestion}, & \text{if } \mathcal{C}_k \neq \emptyset, \\
\texttt{Balanced}, & \text{otherwise},
\end{cases}
\]

The activation of a strategy determines both the type of action to be executed and the selection of destination nodes. When the \texttt{Balanced} strategy is activated, no actions are performed. The available actions are as follows:

\begin{itemize}
  \item \textbf{Consolidation} ($\texttt{consolidate}$): reduces placement cost by migrating VNFs from high-cost nodes to lower-cost nodes when the cost budget is exceeded. This action is prioritized when the \texttt{Cost} strategy is activated.
  
  \item \textbf{Replication} ($\texttt{replicate}$): creates additional VNF instances for applications experiencing degradation or failures. The agent selects affected applications, identifies their dominant user region, and deploys missing VNFs on a suitable destination node. This action is prioritized when the \texttt{Congestion} strategy is activated.

  \item \textbf{Migration} ($\texttt{move}$): relocates VNFs from overloaded nodes. For each overloaded node, candidate VNFs are prioritized based on application latency and moved to a lower-cost, lower-utilization destination. This action is prioritized when the \texttt{Overload} strategy is activated.
\end{itemize}

We summarize the prioritization of placement actions under each strategy in Table~\ref{tab:strategy_actions}. 

\begin{table}[t]
  \small
  \centering
  \caption{Strategy-dependent prioritization of placement actions}
  \label{tab:strategy_actions}
  \begin{tabular}{llp{6cm}}
  \hline
  \textbf{Selected Strategy} & \textbf{Primary Action} & \textbf{Subsequent Actions} \\
  \hline
  \texttt{Cost}       & \texttt{Consolidate} 
              & \texttt{Replicate} (if degraded applications); \texttt{Move} (if overloaded nodes and budget allows) \\
  
  \texttt{Overload}    & \texttt{Move} 
              & \texttt{Replicate} (if degraded applications); \texttt{Consolidate} (if cost constraints apply) \\
  
  \texttt{Congestion}  & \texttt{Replicate} 
              & \texttt{Move} (if overloaded nodes exist); \texttt{Consolidate} (if cost constraints apply) \\
  
  \texttt{Balanced}    & \texttt{None} 
      & \texttt{None}  \\
  \hline
  \end{tabular}
\end{table}

The previous actions select the destination node using the next scoring function:
\[
\mathrm{score}(n) =
d_{\text{users}}(n)
  + \alpha \cdot \mathcal{U}^{\text{n}}_k(n)
  + \beta \cdot \mathcal{P}_{n}
  + \Pi_{\text{region}}(n),
  \]
where:
\begin{itemize}
      \item $d_{\text{users}}(n)$ is the mean distance between node $n$ and the current users of the application,
      \item $\mathcal{U}^{\text{n}}_k(n)$ is the utilization of node $n$ in window $W_k$,
      \item $\mathcal{P}_{n}$ is the cost associated with node $n$,
      \item $\Pi_{\text{region}}(n)$ is a binary region penalty equal to $0$ for nodes in the dominant user region and $\delta_{\text{region}} = 25.0$ otherwise.
\end{itemize}

The relative importance of utilization and cost in the scoring function is controlled by the strategy-dependent penalty coefficients $\alpha$ and $\beta$. In the \texttt{cost} strategy, cost is strongly prioritized ($\beta = 60.0$) over utilization ($\alpha = 25.0$). The \texttt{overload} strategy emphasizes utilization ($\alpha = 55.0$) while reducing the influence of cost ($\beta = 20.0$). The \texttt{congestion} strategy assigns moderate importance to utilization ($\alpha = 25.0$) and a lower weight to cost ($\beta = 12.0$), reflecting its focus on proximity and responsiveness. Finally, the \texttt{balanced} strategy adopts intermediate values ($\alpha = 35.0$, $\beta = 20.0$), providing a compromise between load distribution and cost efficiency.

The total number of actions per window is limited by a fixed budget of $B = 4$ actions, ensuring bounded intervention and preventing excessive oscillations in placement decisions.

To evaluate the adaptive behavior of the system, we introduce a controlled workload perturbation through a \emph{HotspotUsers} event. This event dynamically concentrates demand on a latency-sensitive application by injecting additional users at specific points in time. Concretely, at $t=1000$, 60 additional users of the \emph{Perception Pipeline} application are created at specific \textit{MEC} node, generating a sudden increase in request load. At $t=1800$, these users are relocated to a neighboring node, introducing a spatial shift in demand. Finally, at $t=2600$, 40\% of these users are removed, partially relieving the system. This sequence creates a non-stationary workload characterized by abrupt load increases, spatial redistribution of demand, and partial recovery. These conditions are designed to trigger adaptive behavior in the placement agent, particularly VNF migration (\emph{move}) actions in response to node overload.


\subsection{Results and Analysis}

The comparison between the three placement strategies (random, greedy, and multi-agent) is conducted under the same nominal user workload and the same workload perturbation introduced by the \emph{HotspotUsers} event.  In the random strategy, VNFs are assigned to nodes arbitrarily, and users are associated with applications using a round-robin scheme. Additional users introduced by the hotspot event reuse already deployed VNF instances. In contrast, the greedy strategy deploys application instances on demand: each time a user is created, a new instance of the requested application is generated and all its VNFs are deployed on the \textit{MEC} node where the user is located. As a result, no sharing of VNF replicas occurs, and each user is served by an independent service chain. This behavior is consistently maintained during the hotspot event.

After simulating 5000 time units, we obtain the results shown below. Figure~\ref{fig:response_breakdown_comparison} illustrates the evolution of the response time decomposition (network, waiting, and processing) for each application (rows) and strategy (columns). It is important to note that the time scales differ across strategies. 
The greedy strategy achieves the lowest overall latency, where processing time becomes the dominant component. In contrast, in the random and multi-agent strategies, network latency has a more significant impact. The \emph{Perception Pipeline} application, which is heavily affected by the hotspot, shows a clear performance degradation from $t = 1000$, when the demand spike increases multiple components of the response time. In the random and multi-agent strategies, the increase is mainly reflected in waiting time, whereas in the greedy strategy the network becomes the primary bottleneck due to the concentration of deployments at the edge. Notably, the impact of the hotspot is effectively mitigated only in the multi-agent approach, which dynamically repositions VNFs in response to the changing load conditions. Across all applications, the multi-agent strategy provides a more balanced trade-off between the two baseline approaches, avoiding the extreme behaviors observed in both the random and greedy strategies.

\begin{figure}[t]
  \centering
  \includegraphics[width=0.8\textwidth]{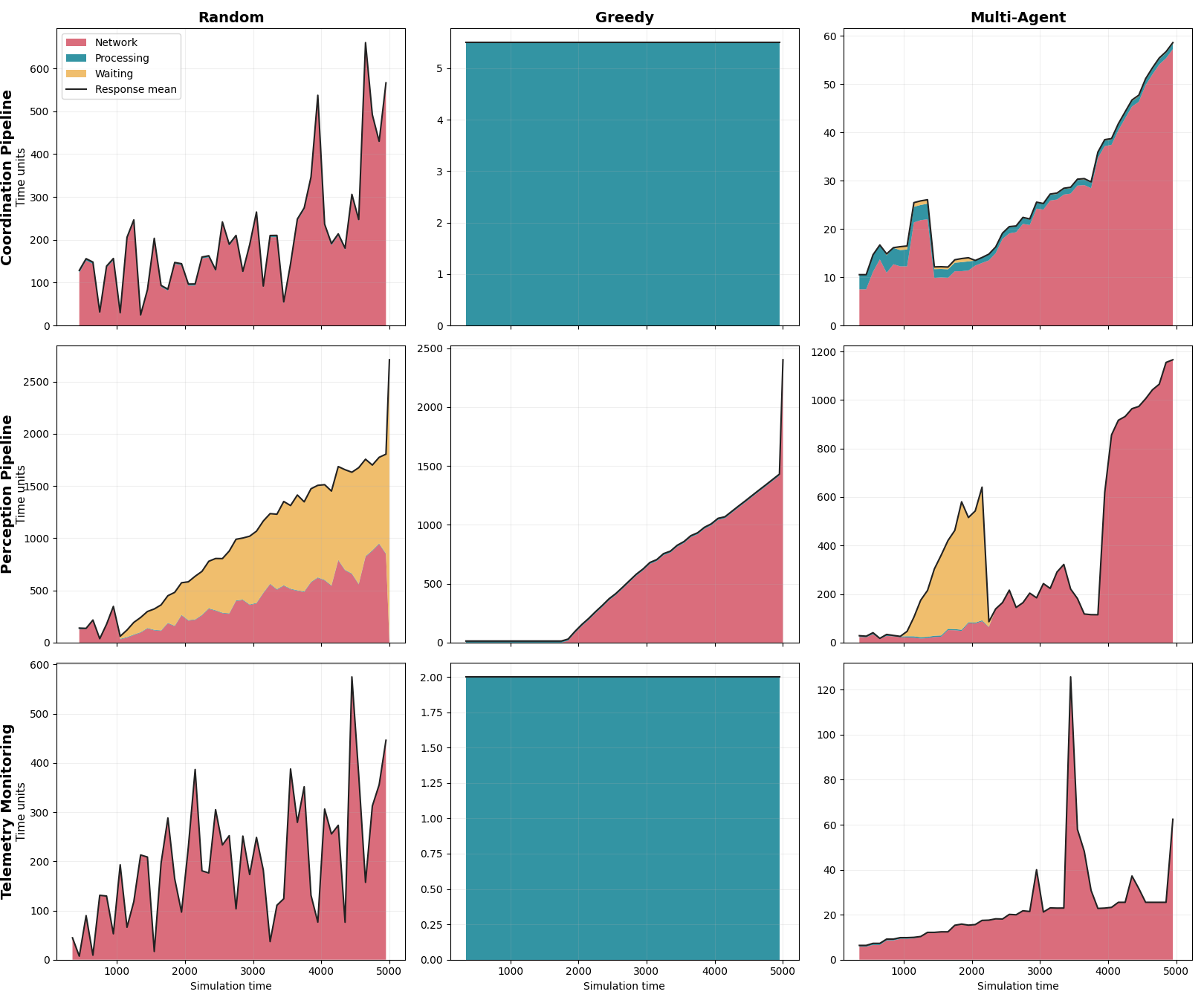}
  \caption{Evolution of agent strategies over time for each application}
  \label{fig:response_breakdown_comparison}
\end{figure}

\begin{figure}[t]
  \centering
  \includegraphics[width=0.8\textwidth]{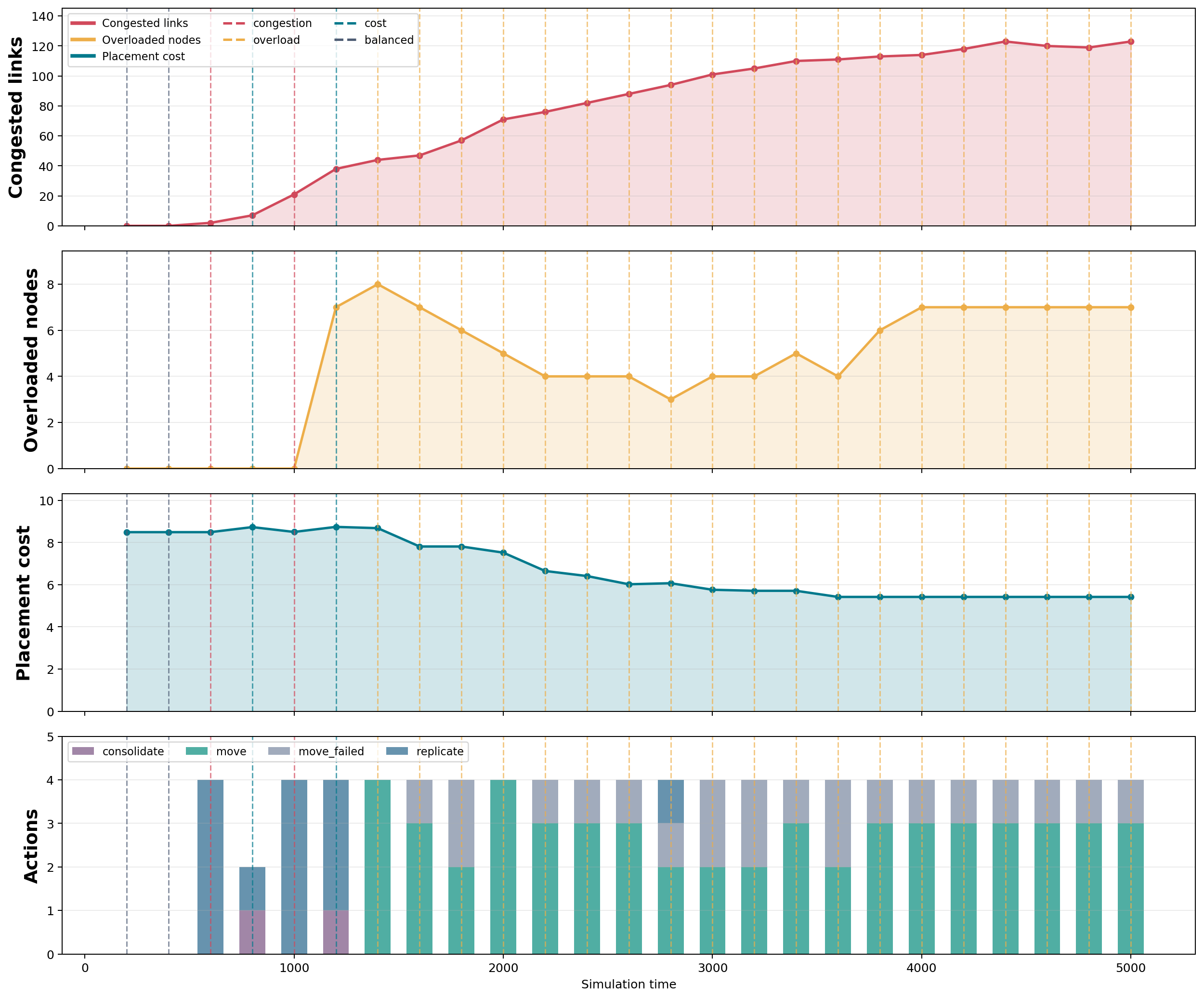}
  \caption{Evolution of multi-agent actions over time}
  \label{fig:multi_agent_strategy_timeline}
\end{figure}

Figure~\ref{fig:multi_agent_strategy_timeline} shows the sequence of actions performed by the multi-agent system across simulation windows. The first three rows present key metrics observed by the \textit{Monitoring Agent}, namely the number of congested links, overloaded nodes, and the aggregated placement cost. The last row displays the actions executed by the \textit{Placement Agent}. Vertical separators indicate window boundaries and also reflect the dominant condition identified by the monitoring agent in each window, which guides the subsequent decisions of the placement agent. The observed actions illustrate how the placement agent attempts to correct the system state over time. Initially, consolidation and replication actions are performed to address cost inefficiencies and link congestion, particularly under increasing load. As the simulation evolves, migration (\emph{move}) actions become dominant, aiming to alleviate node overload by redistributing VNFs across the infrastructure.

Finally, the total number of deployed VNFs highlights the resource footprint of each strategy: approximately 41 for random, 567 for greedy, and 56 for the multi-agent approach. This comparison reflects the trade-off between performance and resource efficiency, where the multi-agent strategy achieves a more balanced deployment.



\subsection{Discussion}

Up to this point, the interaction between agents and the simulation has remained implicit in the description of the experimental scenario. Here, we make this interaction explicit by examining the role of MCP as the abstraction that enables and structures agent-driven control of the simulation.

Our agents operate through a set of standardized tools that expose the capabilities of the simulation. The execution flow at each window can be summarized as follows:

\begin{enumerate}
    \item The simulation advances through control operations such as \texttt{schedule\_for} and \newline \texttt{wait\_un\-til\_ready}.
    \item The \textit{Monitoring Agent} queries system metrics using tools such as \texttt{get\_si\-mula\-tion\_appli\-cation\_me\-trics} and \texttt{get\_simulation\_network\_metrics}.
    \item Based on these observations, the agent constructs the snapshot $S_k$.
    \item The \textit{Placement Agent} consumes $S_k$ and selects a placement strategy.
    \item If required, additional information is retrieved through tools such as \texttt{list\_si\-mu\-la\-tion\_users}, \texttt{list\_si\-mu\-la\-tion\_appli\-cation\_vnfs}, and \texttt{list\_si\-mu\-la\-tion\_no\-de\_pla\-ce\-ments}.
    \item Placement actions are executed via tools such as \texttt{replicate\_appli\-cation\_vnf} and \texttt{move\_appli\-cation\_vnf}.
\end{enumerate}

This interaction pattern highlights that MCP does not represent direct communication between agents, but rather a shared interface that mediates access to the simulation state and operations. Agents remain logically independent and coordinate implicitly through the environment.

From this perspective, MCP fulfills three key roles. First, it provides a \textit{common interface}, allowing agents to interact with the simulation through a consistent set of operations without accessing internal data structures. Second, it enables \textit{decoupling}, as agent logic is expressed in terms of tool invocations and remains independent of the execution context, whether local or remotely exposed through MCP transports. Third, it supports \textit{traceability}, as every interaction is recorded with information about the actor, simulation window, invoked tool, input and output summaries, and execution status.

This separation makes it possible to audit the behavior of the system at a fine-grained level, including what the monitoring agent observed, which strategy the placement agent selected, and how each action was materialized through tool calls.

Overall, MCP transforms the simulation into an interactive and programmable environment, where behavior is defined through sequences of observable and reproducible interactions. This perspective shifts the role of simulation from a static evaluation tool to a dynamic system that can be systematically explored and controlled by external agents.

\section{Conclusions}~\label{sec:conclusions}
In this work, we presented YAIFS, a service-oriented evolution of YAFS that redefines the role of simulation in distributed cloud--edge systems. The proposed framework introduces a layered architecture that progressively exposes different levels of abstraction, from low-level simulation control to high-level orchestration. This design enables both fine-grained modeling flexibility and structured interaction through standardized interfaces.

A central contribution of this work is the integration of the Model Context Protocol (MCP) as a unified interaction layer between external agents and the simulation environment. By exposing simulation capabilities as a set of tools and resources, MCP decouples agent logic from simulator internals and enables a consistent interaction model across different execution contexts. This abstraction allows simulations to be treated as controllable and observable entities, rather than static artifacts.

To demonstrate these capabilities, we presented two complementary scenarios. First, an LLM-based assistant enables users to interact with the simulation through natural language, illustrating how MCP can support intuitive and high-level control. Second, a multi-agent scenario showcases how specialized agents can monitor system conditions and dynamically adapt placement decisions through structured interactions. Together, these scenarios highlight the flexibility of the framework and the role of MCP in enabling agent-driven experimentation.

Overall, this work introduces a shift in perspective by treating simulation as a service that can be programmatically accessed, controlled, and evolved. This approach opens new possibilities for integrating simulation environments within AI-driven workflows, where agents can reason about system behavior, explore alternative configurations, and automate experimental processes.

The YAIFS framework is available at \url{https://github.com/acsicuib/YAIFS}, where we provide the full implementation, documentation, and a collection of representative YAFS scenarios re-implemented using the proposed interaction layers (API and service-oriented interface), as well as the illustrative use cases presented in this work, including the LLM-based interaction and the multi-agent MCP scenario. Additional implementation details and experimental configurations, which are omitted here for brevity, are also provided to facilitate reproducibility and further exploration.

Future work will explore the integration of learning-based agents, including reinforcement learning and advanced LLM-based planning strategies, as well as the coordination of multiple agents operating over shared simulation environments. In addition, extending MCP interoperability with external orchestration tools and real-world deployment frameworks may enable tighter coupling between simulated and operational systems, further bridging the gap between experimentation and deployment.

\section*{Acknowledgments}
This work was supported by Grant PID2024-158637OB-I00, funded by MICIU/AEI/\-10.13039/\-501100011033 and by ``ERDF A way of making Europe'' (ERDF/EU).

\bibliographystyle{elsarticle-num}
\bibliography{sample}

\end{document}